\documentclass[twocolumn, secnumarabic,  floatfix, nofootinbib, nobalancelastpage]{revtex4-2}

\def\TitleOfPaper{How Well Does Kohn-Sham Regularizer Work for Weakly Correlated Systems?}
\def\preprintlinklocation{http://dft.uci.edu + PAPER REF}

\usepackage{amsmath}
\usepackage{physics}
\usepackage{amssymb}
\usepackage{float}
\usepackage[dvipsnames]{xcolor}

\definecolor{TITLECOL}{rgb}{0.05,0.25,0.85}
\definecolor{CONTENTSCOL}{rgb}{0.1,0.2,0.7}
\definecolor{URLCOL}{rgb}{0,0.52,0.83}
\definecolor{LINKCOL}{rgb}{0.05,0.5,0}
\definecolor{CITECOL}{rgb}{0.25,0,0.48}
\definecolor{SECOL}{rgb}{0.07,0.31,0.80}
\definecolor{SSECOL}{rgb}{0.26,0.19,0.75}

\newcommand{\coloredtitle}[1]{\title{\textcolor{TITLECOL}{#1}}}
\newcommand{\coloredauthor}[1]{\author{\textcolor{CITECOL}{#1}}} 

\usepackage{graphicx}

\def\PublicationsLink{http://dft.uci.edu/publications.php}
\def\preprintlink{ \href{\preprintlinklocation}{\TitleOfPaper} }
\def\preprinttext{~}

\usepackage{fancyhdr}
\makeatletter
\fancypagestyle{titlepage}
{	\lhead{\textsc{\preprinttext\ ~ \href{\PublicationsLink}{~}}}
	\chead{}
	\rhead{\preprintlink}
	\lfoot{}}
\makeatother
\def\preprintlink{ 
	\href{\preprintlinklocation}
        {
~}
	}

\definecolor{Green}{rgb}{0.016,0.627,0}
\definecolor{Plum}{rgb}{0.17,0,0.45}
\definecolor{LBlue}{rgb}{0,0.34,0.45}
\definecolor{Sepia}{rgb}{0.37,0.17,0.02}
\definecolor{BurntOrange}{rgb}{0.78,0.39,0}

\usepackage{hyperref}
\hypersetup{
    breaklinks=true,
    bookmarksopen=true,
    bookmarksnumbered=true,
    colorlinks=true,
    linkcolor=LINKCOL,
    linktocpage=true,
    citecolor=CITECOL,
    filecolor=magenta,      
    urlcolor=URLCOL,
}

\usepackage{verbatim}

\def\bea{\begin{eqnarray}}
\def\eea{\end{eqnarray}}
\def\ben{\begin{equation}}
\def\een{\end{equation}}
\def\benu{\begin{enumerate}}
\def\enu{\end{enumerate}}

\def\bei{\begin{itemize}}
\def\eei{\end{itemize}}
\def\beit{\begin{itemize}}
\def\eit{\end{itemize}}
\def\benu{\begin{enumerate}}
\def\enu{\end{enumerate}}

\def\sss{\scriptscriptstyle\rm}

\def\1var{(\bx_1...\bx\N)}

\def\bx{{x}}

\def\s{_{\sss S}}
\def\xc{_{\sss XC}}

\def\N{_{\sss N}}
\def\H{_{\sss H}}

\def\sph_int{ {\int d^3 r}}

\graphicspath{{figures/}}

\usepackage[caption = false]{subfig}
\usepackage{booktabs}
\usepackage[normalem]{ulem}
\usepackage[encapsulated]{CJK}

\begin{document}
\begin{CJK*}{UTF8}{gbsn}
\sf
\coloredtitle{\TitleOfPaper}



\coloredauthor{Bhupalee Kalita}
 \affiliation{Department of Chemistry, University of California, Irvine}

\coloredauthor{Ryan Pederson}%
\affiliation{Department of Physics and Astronomy, University of California, Irvine}%
\coloredauthor{Jielun Chen (陈捷伦)}%
\affiliation{Department of Physics and Astronomy, University of California, Irvine}%
\coloredauthor{Li Li (李力)}%
\affiliation{%
Google Research
}%
\coloredauthor{Kieron Burke}%
\affiliation{Department of Chemistry, University of California, Irvine}%
\affiliation{Department of Physics and Astronomy, University of California, Irvine}

\begin{abstract}
 Kohn-Sham regularizer (KSR) is a differentiable machine learning approach to finding the exchange-correlation functional in Kohn-Sham density functional theory (DFT) that works for strongly correlated systems. Here we test KSR for weak correlation. We propose spin-adapted KSR (sKSR) with trainable local, semilocal, and nonlocal approximations found by minimizing density and total energy loss. We assess the atoms-to-molecules generalizability by training on one-dimensional (1D) H, He, Li, Be, Be$^{++}$ and testing on 1D hydrogen chains, LiH, BeH$_2$, and helium hydride complexes. The generalization error from our semilocal approximation is comparable to other differentiable approaches, but our nonlocal functional outperforms any existing machine learning functionals, predicting ground-state energies of test systems with a mean absolute error of 2.7 milli-Hartrees.

 \end{abstract}

\maketitle
\end{CJK*}

Determining the ground-state properties of many-electron systems is fundamental to molecular modeling problems in chemical and material sciences. However, solving the Schr\"odinger equation explicitly for more than a few hundred electrons is computationally intractable. Among several methods of approximation, Kohn-Sham density functional theory (KS-DFT or simply DFT)~\cite{HK64, KS65}, a method based on the electron density distribution rather than the many-electron wave function, provides chemically useful results with $O(N^3)$ scaling for an $N$-electron system~\cite{DG90}. DFT is formally exact, but the exchange-correlation (XC) energy, resulting from the quantum-mechanical interaction between electrons, must be approximated in practice. Hundreds of XC energy functional approximations have been formulated in the past few decades~\cite{mardirossian2017thirty}. Functionals can be designed non-empirically, for example using physics and chemical-based intuition and satisfying known exact constraints~\cite{PBE96}, or can involve some fitting to reference data~\cite{ZT06}. However, in any approach, these functional approximations do not yield chemical accuracy in general, that is, with errors less than 1.6 milli-Hartrees (mH) in atomic units (or 1 kcal/mol). Improving the accuracy of XC functional approximations often incurs additional computational cost in the practical DFT calculation~\cite{B12}. However, there is no systematic way in general to develop and improve XC functional approximations.

In recent years, machine learning (ML) has been used to find better DFT approximations. Attempts have been made to enhance either the speed or accuracy of DFT. Some used ML techniques to boost computational efficiency by approximating the non-interacting kinetic energy without solving the KS equations~\cite{SRHB12, LSPB16, BVLM17, KLMB21}. In an effort to improve the accuracy of ML-DFT, a significant leap was achieved by Nagai et al.~\cite{NAS20}, who used a neural network (NN) model to approximate the XC functional and trained it with high accuracy coupled cluster (CCSD(T)) energies and densities of just three small molecules, while self-consistently solving the KS equations. This functional impressively generalized to 148 small molecules~\cite{CRTP91} to predict their energies and densities with accuracies comparable to human-designed functionals. However, the test set atomization energies were not chemically accurate. Also, they didn't have access to gradient information and were therefore limited to a gradient-free optimization scheme, which is inherently slow, often suffers poor convergence issues, and is difficult to scale to more complex NN models.

In DFT, many useful properties are extracted from the density, although an XC functional approximation need not produce accurate densities along with accurate energies~\cite{KSB13}. In KS-DFT, we calculate the density self-consistently, and there is a nonlinear dependence of the XC functional on the density. Learning this relationship requires not only the ground truth mapping of the functional inputs to outputs but also how the functional performs in the underlying process. Hence the use of differentiable programming~\cite{BPRS17} becomes more intuitive~\cite{IEFT19}. With differentiable programming, conditioning the networks with physical insights becomes much simpler, and it can further help to ease the process of training.

Recently, Li et al.~\cite{LHPB21} made a valuable step in this direction by considering the entire DFT self-consistent calculation as a differentiable program. They implemented an end-to-end differentiable DFT code for 1-dimensional (1D) systems using JAX~\cite{jax2018github}, a library that provides differentiation, vectorization, just-in-time compilation, and other composable transformations of Python and NumPy programs~\cite{HMVO20}. They parameterized the XC functional with an NN which incorporated non-local information about the density, along with known physical constraints. The self-consistent KS calculations were embedded into the training process by backpropagating the gradients through the KS iterations. It was dubbed the Kohn-Sham regularizer (KSR). It could yield chemically accurate energies for uniformly separated 1D hydrogen chains at any separation by training on highly accurate energies and densities from only a few separations.

Following a similar approach, Kasim and Vinko~\cite{KV21} implemented an end-to-end differentiable DFT code in 3D for Gaussian-type orbitals and trained local and semi-local NN-based XC functional approximations, evaluating performance on small molecules. In another work, Dick et al.~\cite{DS21} constructed a semilocal XC functional that was carefully curated to account for several known exact conditions and pretrained to match SCAN, a popular meta-GGA functional~\cite{SRP15}. While both of these works explore the generalizability of ML approximations for weakly correlated molecules with differentiable DFT codes, they do not incorporate global information, and their accuracy is limited to that of human-designed semilocal functionals. A slightly different approach involves introducing an ML correction term to a nonempirical or semi-empirical XC functional within a KS-DFT self-consistent framework~\cite{DF-S20, CZWE21}. In such an approach, only a portion of the XC energy is approximated using ML and the functionals retain the characteristics of the baseline XC functional used. 
The recently proposed ML local hybrid functional, DM21~\cite{DM21}, addresses spin-symmetry breaking and delocalization error in DFT functionals. Consequently, it performs well on several main-group benchmark datasets and also correctly dissociates molecules. Unlike KSR, this functional is trained on large datasets of highly accurate reaction energies (not densities) in the loss function without explicitly supervising the self-consistent iterations.

[C3] Li et al.~\cite{LHPB21} explored the generalizability of KSR for a few strongly correlated systems with stretched bonds which is a completely different domain from most chemical applications of DFT. The aim there was to generate accurate binding energy curves (all the way to the dissociation limit) using the
entire density (for the nonlocal approximation called global-KSR), using inputs at only two separations, for unpolarized hydrogen chains.   The generalizability was
in finding the entire bond-dissociation energy curve of these chains. Moreover, only the total density was used and not the spin densities. 

In the present work,  we propose spin-polarized versions of local, semilocal, and nonlocal XC functional approximations within a differentiable spin-DFT implementation of KSR. We modify these approximations to predict XC energy densities using spin-densities as feature vectors while optimizing the NN parameters using total density and energy loss. Contrary to Ref.~\cite{LHPB21}, we test the KSR approach in the domain of routine DFT calculations in chemistry, namely in and
around equilibrium bond lengths. We find the remarkable result that training on energies and densities of a few atoms (and ions) alone produces accurate ground-state energies for equilibrium molecules
(very reminiscent of the use of appropriate norms while avoiding using any covalent bond energies).
We train and test on a variety of different elements, to obtain the generalizability relevant to chemistry.
Almost all previous work in the chemical domain tests various approximate functional forms employing the
standard ingredients locally~\cite{NAS20, DS21, KV21}.  Our work achieves high accuracy
using the total density and is not limited to a specific set of human-chosen features.
 
The practical implementation of DFT involves solving the Kohn-Sham (KS) equations to calculate the ground-state electron density,
\begin{equation}
    \left\{-\frac{1}{2}\nabla^2 + v\s[n](\mathbf r)\right\}\phi_i (\mathbf r) = \epsilon_i\phi_i(\mathbf r).
    \label{eq:KS}
\end{equation}
The electron density, $n(\mathbf r)$, is the sum of the probability density over all occupied one-electron KS orbitals, $n(\mathbf r) = \sum_i|\phi_i(\mathbf r)|^2$. The KS potential, $v\s[n](\mathbf r)$, contains the external one-body potential, the Hartree potential, and the XC potentials,
\begin{equation}
    v\s[n](\mathbf r)= v(\mathbf r) + v\H[n](\mathbf r) + v\xc[n] (\mathbf r).
\end{equation}
The XC potential is the functional derivative of the XC energy, $E\xc[n]$, with respect to the electron density~\cite{KS65}, $v\xc[n](\mathbf r) = \delta E\xc[n](\mathbf r)/\delta n(\mathbf r)$. We can express $E\xc[n]$ in terms of an XC energy density per electron, $\epsilon\xc[n](\mathbf r)$:
\begin{equation}
  E\xc[n] = \int d^3r \, \epsilon\xc[n](\mathbf r) \, n(\mathbf r). 
\end{equation}
 The ground-state energy is calculated from the self-consistent density by summing the non-interacting kinetic energy, $T\s$, the external potential energy, $V$, the Hartree energy, $U$, and the XC energy,
\begin{equation}
    E_0 = T\s[n] + V[n] + U[n] + E\xc[n].
\end{equation}
The computational efficiency is also affected by the level of approximation used for the XC functional~\cite{PRTC05}.

Density matrix renormalization group (DMRG)~\cite{White92} can be used to efficiently generate highly accurate benchmark energies and densities for these 1D analog systems. We can address such systems using 1D KS-DFT calculations as well with suitable XC energy functional approximations, such as the 1D local spin-density approximation (LSDA) which was constructed in Ref.~\citenum{BSWW15} from the 1D exponentially repelling uniform electron gas.

In essence, KSR is a ML-DFT regularization technique that utilizes a differentiable analog of the standard self-consistent DFT computational flow during training to train a suitable parameterized model for $E\xc[n] = {E\xc}_{,\theta}[n]$, where $\theta$ are trainable parameters~\cite{LHPB21}. In this work, we consider NN-based (neural) XC models, but KSR as a regularization technique can apply more broadly to any differentiable model choice. 
Knowledge of physical properties and constraints in the exact XC functional can help guide the construction of a neural XC approximation. The NN that parameterizes the XC functional in KSR is carefully curated to account for a few of the expected behaviors of the exact XC functional.  Nonlocality is facilitated by adding a global convolution layer in ${\epsilon\xc}_{,\theta}[n]$ to help capture long-range interactions. The sigmoid linear unit (SiLU or Swish)~\cite{SEK18, PBQ18} activation function is used throughout because of its infinite differentiability. The KSR network is also complemented with a self-interaction gate (SIG) that partially cancels the self-interaction error by mixing in a portion of  Hartree energy density to $\epsilon\xc$. 

In Ref.~\citenum{LHPB21} several neural XC functional models were proposed: a local functional which only depends on the density at each point (KSR-LDA), a semi-local functional that uses local and gradient information about each point (KSR-GGA), and a global functional which utilized the global convolution layer and the SIG described above (KSR-global).

 A main deficiency of the KSR technique in Ref.~\citenum{LHPB21} is that it does not explicitly account for spin, and so may not generalize well for spin-polarized systems. Extending this technique and associated NN models to spin DFT requires a differentiable framework that can backpropagate through resulting spin densities. 
Spin is often incorporated in the neural XC functional using relative polarization, $\zeta$, as a feature~\cite{NAS20}. For up and down spin densities,  $\{n_{\uparrow}$, $n_{\downarrow}\}$, $\zeta = (n_{\uparrow}-n_{\downarrow})/n$. While $\zeta$ can be introduced as an additional input channel to KSR neural $\epsilon\xc$, its scale can be very different relative to $n$ in general. Instead, we use up and down spin densities as input features, which have similar scales. The usual models and concepts for KSR can be extended to obtain a spin-adapted KSR (sKSR).

\begin{figure*}[ht!]
     \centering
     \subfloat[]{\includegraphics[width=0.51\textwidth]{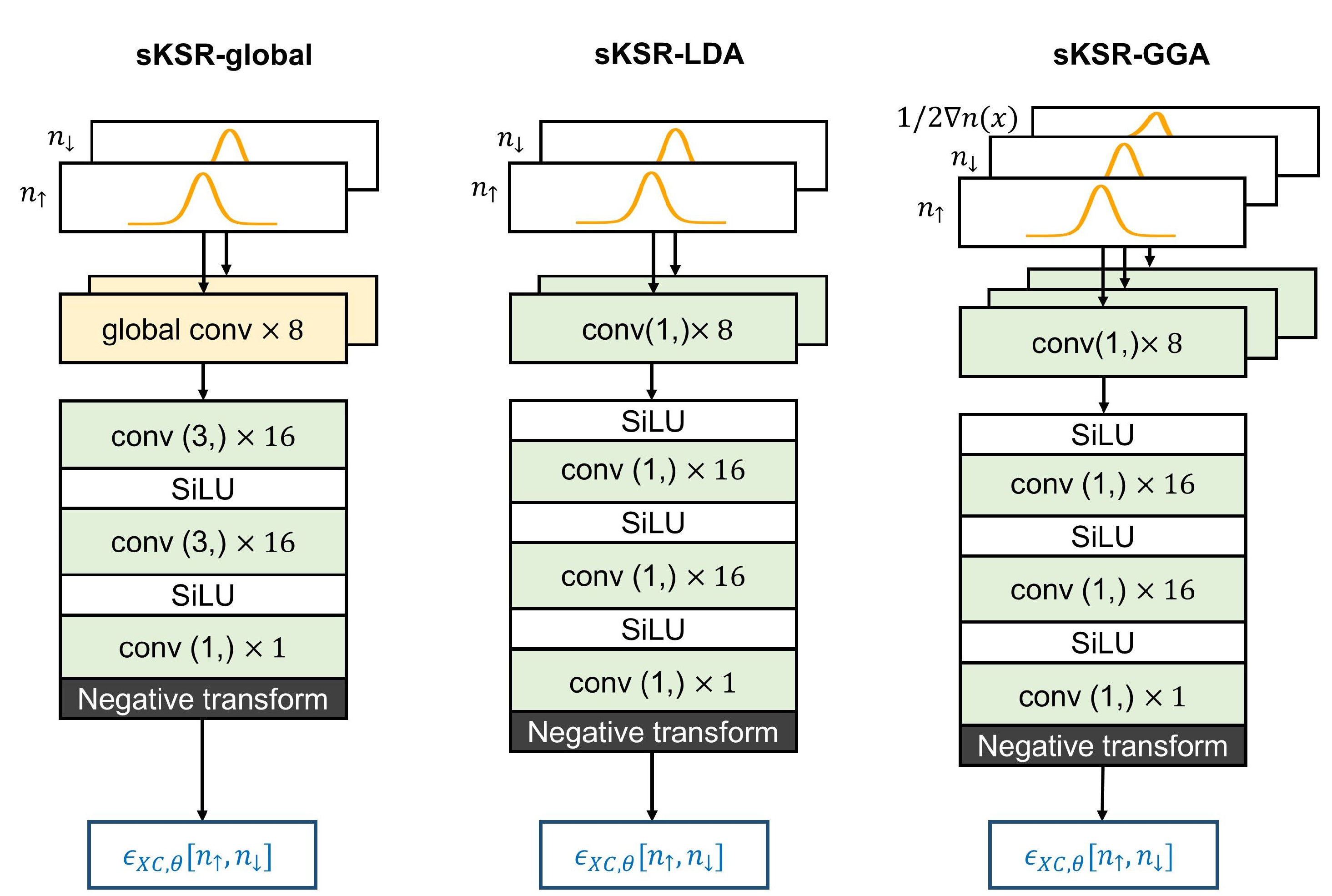}\label{fig:a}}
    \hfill
    \subfloat[]{\includegraphics[width=0.47\textwidth]{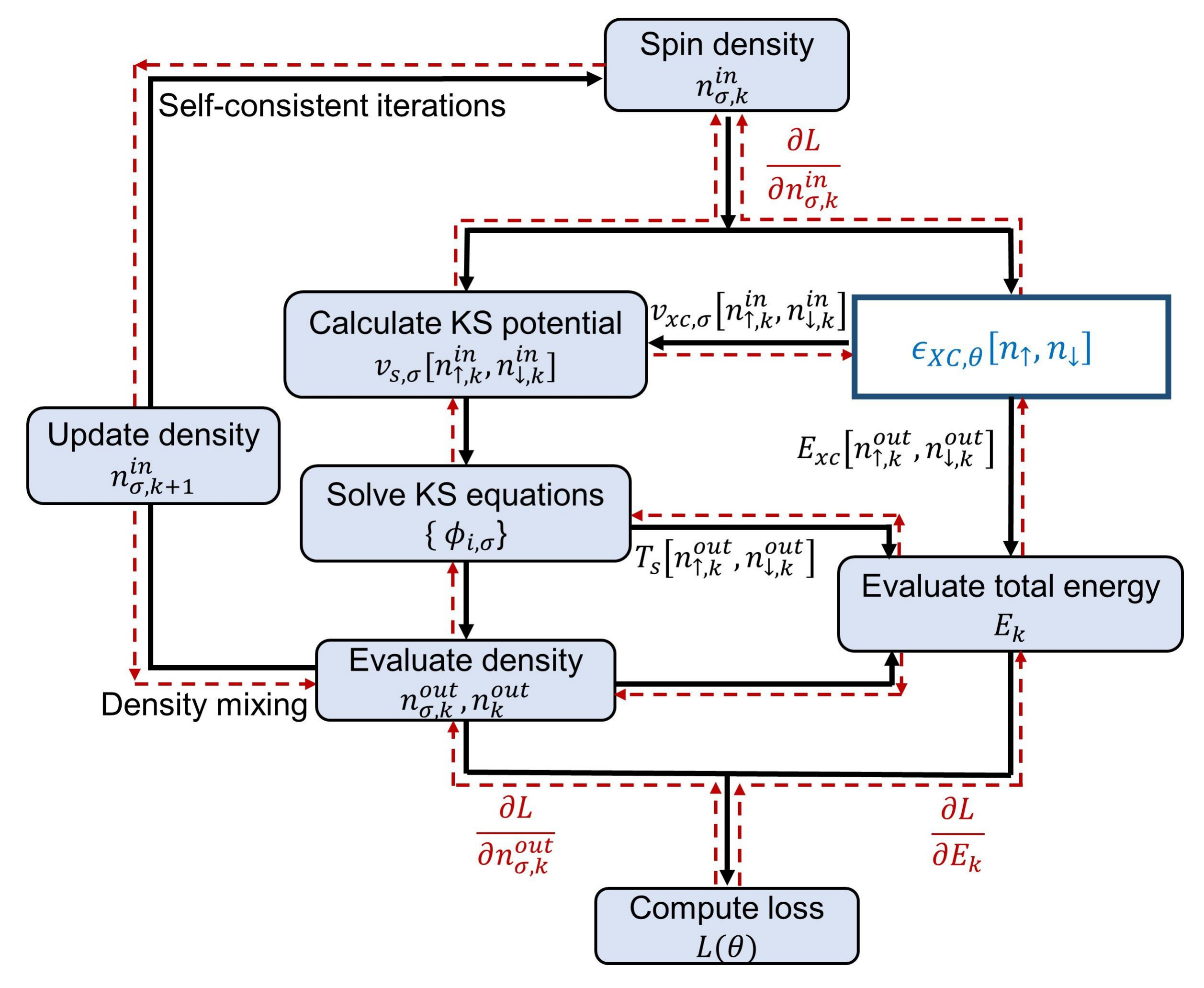}\label{fig:b}}
   \caption{(a) sKSR-global, sKSR-LDA and sKSR-GGA architectures to calculate $\epsilon\xc$ from spin-densities. (b) sKSR -- differentiable KS-DFT with spin-polarization. Black arrows refer to the conventional computational flow. The gradients flow along red-dashed arrows to minimize the loss during training. }
    \label{fig:KSR_LDA_GGA_o}
\end{figure*}

 In sKSR-global, we have a global convolution layer that takes spin densities as inputs, and the kernel takes the form:
\begin{equation}
G\left(n_{\sigma}(x),\xi_p\right) = \frac{1}{2} \xi_p\int dx'n_{\sigma}(x')e^{-|x-x'|/\xi_p},
\end{equation}
where $\sigma\in\{\uparrow, \downarrow\}$ and $\xi_p$ is a trainable parameter that represents an interaction scale.  To keep the number of parameters comparable with KSR-global, we input each spin density to a global convolution layer consisting of $8$ channels. We then concatenate the output on the channel dimension and input it to the latter convolution layers.  For weakly correlated systems and greater generalizability, this approximation does not include any SIG.  The rest of the network architecture is kept unchanged. sKSR-LDA and sKSR-GGA approximations to XC are devoid of global information. For sKSR-LDA, two convolution layers with filter size one and 8 channels map the spin-density to $\epsilon\xc$ at the same spatial point $x$. In sKSR-GGA, we specify the total density gradient explicitly as an additional input channel along with the spin-densities. Instead of using one convolution layer with filter size three, we use three convolution layers with filter size one and 8 channels each. The rest of the sKSR-LDA and sKSR-GGA architectures are also similar to KSR-LDA and KSR-GGA. Fig.~\ref{fig:KSR_LDA_GGA_o}(a) shows the comparative network structures for all three types of approximations. In all cases, the resulting $\epsilon\xc$ is symmetrized with respect to the input of the up and down densities:
\begin{equation}
    \epsilon\xc^{\text{symm}}[n_{\uparrow}, n_{\downarrow}] = \frac{1}{2} \bigg[ \epsilon\xc[n_{\uparrow}, n_{\downarrow}] +  \epsilon\xc[n_{\downarrow}, n_{\uparrow}] \bigg].
\end{equation}

Our approximation replaces the $\epsilon\xc$ in a spin-polarized self-consistent KS-DFT framework.  For spin-polarized systems we perform the above spin-unrestricted KS-DFT procedure, however for unpolarized systems we use spin-restricted KS-DFT to preserve spin-symmetry. Fig.~\ref{fig:KSR_LDA_GGA_o}(b) shows the conventional computational flow and the flow of the gradients during the self-consistent optimization. To train the neural XC functional, we use the following loss function:
\begin{eqnarray}
   L(\theta)= && \underbrace{\mathbb{E}_\mathrm{train}\left[(E^{\text{sKSR}} - E^{\text{DMRG}})^2/N_e\right]}_{\mathrm{energy}\, \mathrm{loss}\, L_E} \nonumber \\
     && +\underbrace{\mathbb{E}_\mathrm{train}\left[
\int dx \, (n^{\text{sKSR}} - n^{\text{DMRG}})^2/N_e\right]}_{\mathrm{density}\, \mathrm{loss}\, L_n},
\label{eq:loss_fn}
\end{eqnarray}
where $E^{\text{sKSR}}$ and $n^{\text{sKSR}}$ are the converged total energy and total density obtained from the neural XC functional approximations, and $E^{\text{DMRG}}$ and $n^{\text{DMRG}}$ are the exact ground-state electronic energy and total density for each of the test systems. The total loss is evaluated as an expectation over training examples, where $N_e$ is the number of electrons for a given training example. All quantities are in atomic units. We only consider the converged energy in the energy loss term rather than the energy trajectory throughout KS iterations, which was explored in Ref.~\citenum{LHPB21}. In this work we find that the self-consistent calculations converge quickly for the small atoms and ions used in training, and incorporating energy loss from each KS iteration minimally affects the efficiency of the optimization process. The gradients are calculated based on the total loss with respect to the parameters, $\theta$, through automatic differentiation. They are back-propagated across the self-consistent cycles and the parameters of the neural XC functionals are updated until the total loss is minimized.

\begin{figure}[htp!]
    \centering
    \includegraphics[width= 0.5\textwidth]{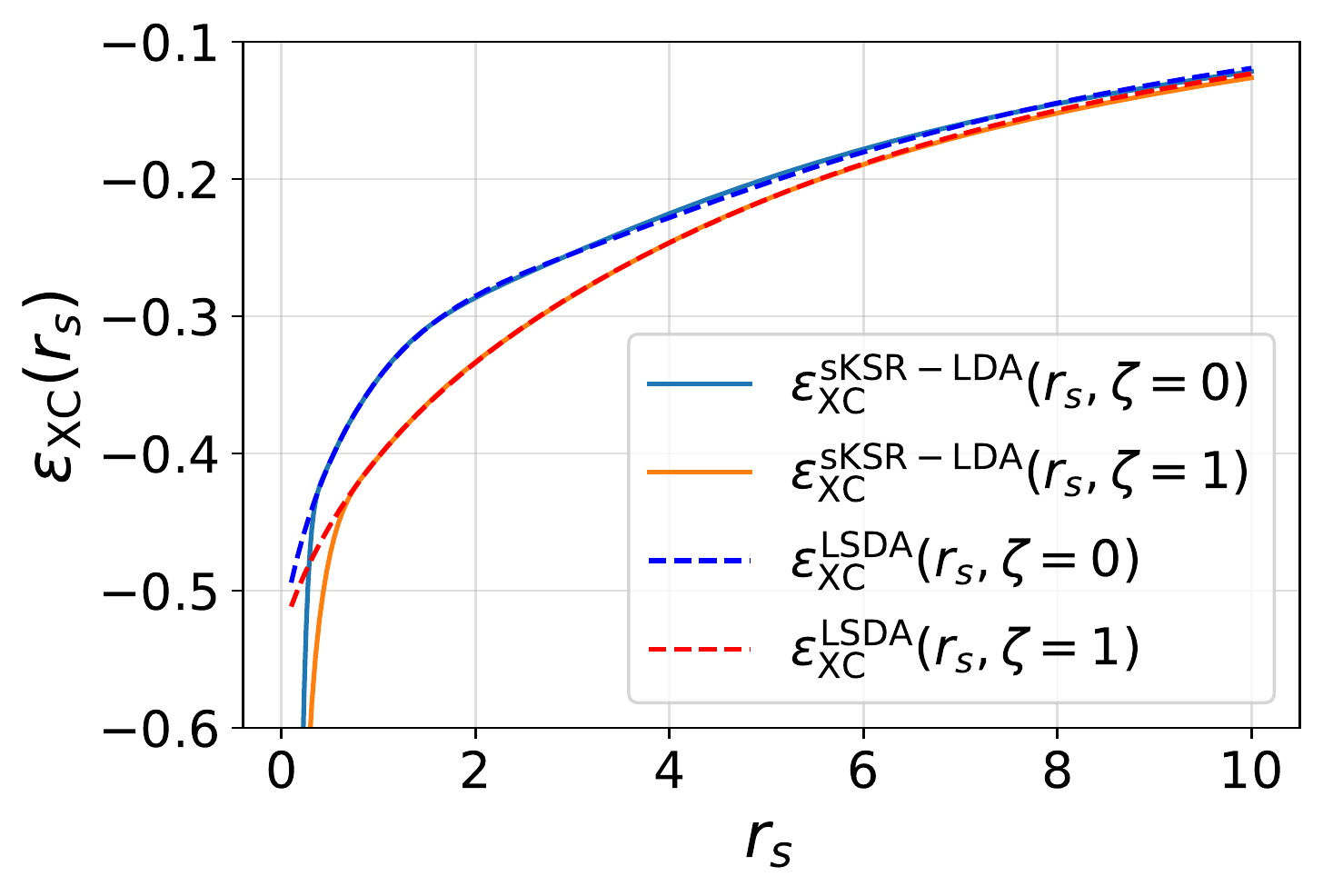}
    \caption{sKSR-LDA trained on 1D LSDA-calculated Li$^{++}$ and He energies and densities. Here $r\s = 1/2n$ and $\epsilon^{\text{unif}}\xc$ corresponds to the XC energy density of the 1D uniform electron gas ~\cite{BSWW15}.}
   \label{fig: ksr_lsda_unif_xc_energies}
\end{figure}

As a simple consistency test, we pose the question: can KSR learn human-designed functionals from their observable results? Here we specifically investigate whether sKSR-LDA can learn the relatively simple but general human-designed 1D LSDA XC functional. Since our sKSR-LDA model utilizes hundreds of parameters, it is unclear whether training on just a few LSDA generated DFT results will yield a neural XC model that matches LSDA. We find that by training sKSR-LDA on LSDA-generated He and Li$^{++}$, we recover the LSDA XC functional almost exactly for unpolarized and fully polarized systems, see Figure~\ref{fig: ksr_lsda_unif_xc_energies}. The sKSR-LDA model deviates at the high-density limit (low $r\s$ limit) due to the limitation that our training densities only consist of $r\s > 0.5$.

\begin{table}[htp!]
    \caption{\label{tab:datasets}Training, validation and test sets for generalizability experiment. The molecules in the test set refer to the relaxed structures.}
  \begin{tabular}{lll}
  \hline
    \textbf{Training}  &    \textbf{Validation}  &   \textbf{Testing}   \\
    \hline
      H, He, Li        &   Be$^+$    &  H$_2$, H$_3$, H$_4$, H$_2^+$, H$_3^+$  \\
        Be, Be$^{++}$ &              &   LiH, BeH$_2$, HeH$^+$ \\
                      &              &    H-He-He-H$^{2+}$ \\
                      &              &   He-H-H-He$^{2+}$ \\
   \hline
  \end{tabular}
\end{table}

Next, we assess generalizability by training sKSR models using a few 1D atomic systems and testing on unseen 1D molecular systems. We trained all three models on DMRG energies and densities of H, He, Li, Be, and Be$^{++}$ and validated on Be$^+$. For training and validation details, see Supporting Information. The trained model was later used to calculate the properties of several molecules in their equilibrium ground-state or relaxed form (see Table~\ref{tab:datasets}). The errors in total energies, ionization, and atomization energies, as well as the average density losses for all three neural XC functional approximations, are reported in Table~\ref{tab:error_table}. Compared to LSDA, the mean absolute error (MAE) in sKSR-LDA calculated energies is reduced by a factor of three. On the other hand, sKSR-global is an order of magnitude higher in accuracy and yields total energies with an MAE of 2.7 mH, not so far from the chemical accuracy limit of 1.6 mH. The cumulative MAEs for the training, validation and test datasets are reported in Supporting Information. 
\begin{table*}[ht!]
\caption{Total energy errors (in mH), density losses (in $10^{-4}$ Bohr$^{-1}$), and errors in ionization potentials for atoms and atomization energies in molecules (in mH) calculated using uniform gas LSDA~\cite{BSWW15}, sKSR-LDA, sKSR-GGA, and sKSR-global respectively, for the training, validation, and test sets in Table~\ref{tab:datasets}.}
 \fontsize{9}{9}\selectfont
  \begin{tabular}{llcccccccccccc}
  \hline
\textbf{Dataset} & \textbf{Symbol} & \multicolumn{3}{c}{\textbf{LSDA}} & \multicolumn{3}{c}{\textbf{sKSR-LDA}} & \multicolumn{3}{c}{\textbf{sKSR-GGA}} & \multicolumn{3}{c}{\textbf{sKSR-global}} \\

\cmidrule{3-5}\cmidrule{6-8}\cmidrule{9-11}\cmidrule{12-14}
&   &  $\Delta E$  & $L_n$ & $\Delta IP$ & $\Delta E$  & $L_n$ & $\Delta IP$ & $\Delta E$  & $L_n$ & $\Delta IP$ & $\Delta E$  & $L_n$ & $\Delta IP$ \\

 \hline
Training & H & 26.6 & 5.35 &  -26.6 & 4.51 & 0.55 & -4.50 & 4.49 & 0.31 & -4.49 & 0.85 & 0.33 & -0.85\\
  & He & 41.4 & 2.89 & -8.46 & 20.2 & 0.63 &  -21.3 & 7.49 & 0.24 & -10.2 & -0.69 & 0.03 & 0.62\\
&  Li & 33.7 & 5.02 & 16.6 & -11.5 & 0.40 & 37.4 & -12.0 & 1.37 & 20.2 & -2.37 & 0.12 & 2.79\\
&  Be & 24.5 & 1.18 & 21.4 & -23.5 & 1.03 & 12.1 & -2.70 & 0.65 & -5.29 & 1.16 & 0.07 & -1.23\\
&  Be$^{++}$ & 55.3 & 0.75 & -18.1 & 29.2 & 0.16 & -46.1 & 6.55 & 0.49 & -34.1 & 0.41 & 0.02 & -1.43 \\\cmidrule{2-14}
& MAE & 36.3  & 3.04 & 18.3 & 17.8 & 0.56 & 24.3 & 6.65 & 0.16 &  14.8 & 1.10 & 0.12 & 1.38\\
\hline
Validation & Be$^+$ & 46.0 & 1.95  & 9.37 & -11.3 & 0.12 & 40.5 & -7.99 & 0.61 & 14.5 & -0.07 & 0.03 & 0.49 \\
& & & & $\Delta AE$ & & & $\Delta AE$ & & & $\Delta AE$ & & & $\Delta AE$ \\
\hline
Test & H$_2$ & 34.04 & 1.82 & 19.2 & 19.5 & 0.35 & -10.5 & 6.83 & 1.99 & 2.14 & -0.73 & 0.07 & 2.43\\
& H$_3$ & 35.6 & 1.93 & 44.3 & 0.45 & 0.21 & 13.1 & -3.07 & 5.57 & 16.5 &  -3.56 & 3.22 & 6.11 \\
& H$_4$ & 32.3 & 3.82 & 74.3 & 7.66 & 1.59 & 10.4 & -9.34 & 4.18 & 27.3 & 2.87 & 1.46 & 0.53\\
& H$_2^+$ & 19.6 & 6.68 & 7.09 & 2.78 & 0.71 & 1.73 & 1.68 & 1.71 & 2.81 & -1.94 & 1.04 & 2.79\\
& H$_3^{+}$ & 31.2 & 0.78 & 22.1 & 20.6 & 1.87 & -11.6 & 15.4 & 11.5 & -6.44 & -0.40 & 0.47 & 2.09\\
& LiH & 30.9 & 3.72 & 29.5 & -8.55 & 2.47 & 1.53 & -16.6 & 3.86  & 9.14 & -4.38 & 0.66 & 2.86\\
& BeH$_2$ & 32.8 & 7.49 & 45.0 & -27.8 & 5.5 & 13.4 & -34.6 & 3.09 & 40.9 & -5.07 & 1.29 & 7.93\\
& HeH$^{+}$ & 37.3 & 1.71 & 4.18 & 18.8 & 0.17 & 1.40 & 5.18 & 0.59  & 2.31 & -1.60 & 0.13 & 0.91\\
& H-He-He-H$^{2+}$ & 36.7 & 14.7 & 46.1 & 5.00 & 6.00 & 35.5 & -9.04 & 2.50 & 24.0 & 5.39 & 4.52 & -6.77\\
& He-H-H-He$^{2+}$ & 46.1 & 7.40 & 36.7 & 19.9 & 6.48 & 20.6 & 4.35 & 4.75 & 10.6 & 0.79 & 5.47 & -2.18\\\cmidrule{2-14}

& MAE & 33.6 &  5.00 & 32.9 & 13.1 & 2.53 & 12.0 & 10.6 & 3.98 & 14.2 & 2.67 & 1.83 & 3.46\\
\hline
  \end{tabular}
\label{tab:error_table}
\end{table*}

 The importance of spin in sKSR can be seen by comparing results with the original 
KSR-global model from Ref.~\cite{LHPB21}. For a valid comparison, we consider KSR-global
without the SIG and train it with the sets from Table.~\ref{tab:datasets}, without
adding the energy trajectory loss. The MAE in KSR-global predictions for total energies
of the test molecules is 10.02 mH, comparable to sKSR-GGA, but much worse than sKSR-global (see Table.~A2 in Supporting Information).
sKSR-global also converges more quickly than KSR-global, reaching lower training losses
with fewer training steps (see Fig.~A7 in Supporting Information). 

The size of our dataset is practically limited by the chemical space provided by 1D and the associated exponential interaction. Even though we are dealing with a much smaller dataset, we trained the sKSR models on the ground-state energies and densities of $5$ atomic systems only and did not include any molecules, contrary to results in Ref.~\citenum{NAS20} and Ref.~\citenum{KV21} which train on derived quantities, such as atomization and ionization energies, and include molecules in training. 

\begin{figure*}[htp!]
     \centering
     \subfloat[]{\includegraphics[width=0.85\textwidth]{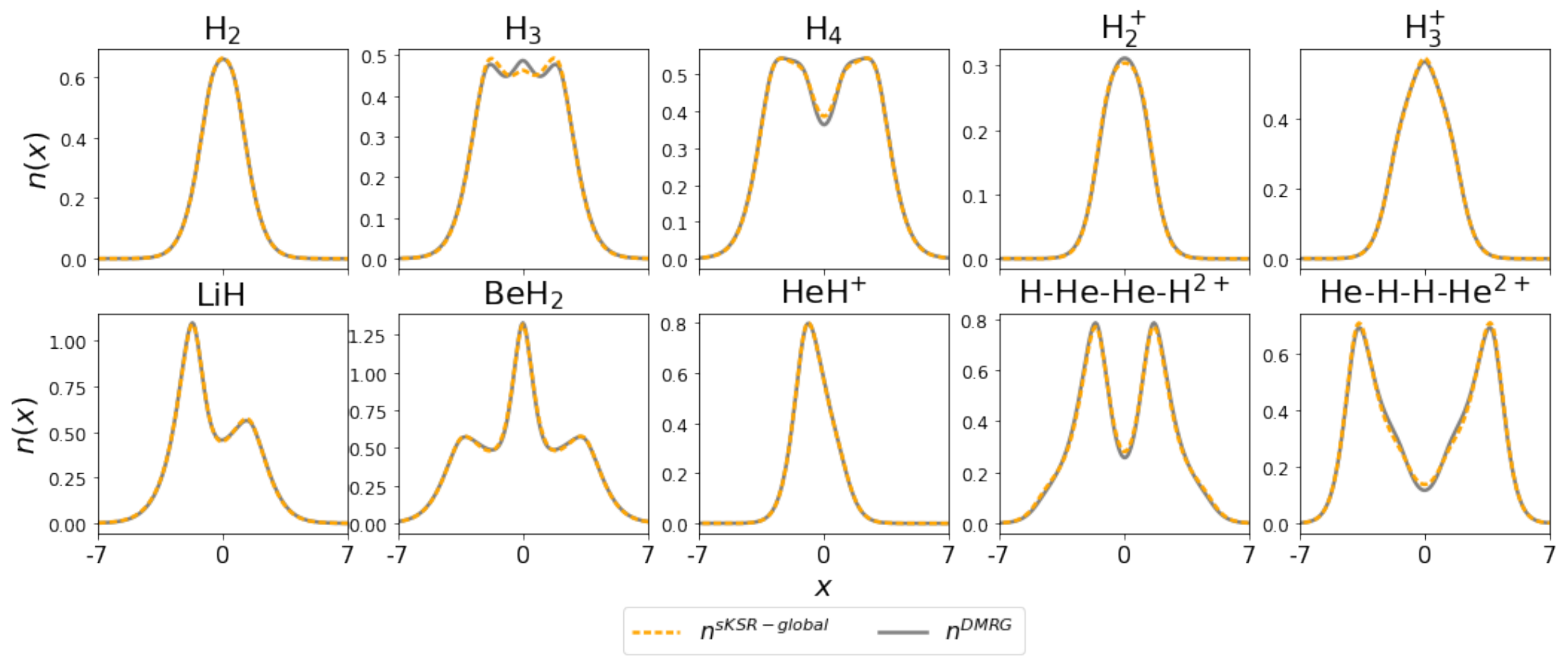}\label{fig:f}}
   \hfill
   \hspace{0.5em}
    \subfloat[]{\includegraphics[width=0.85\textwidth]{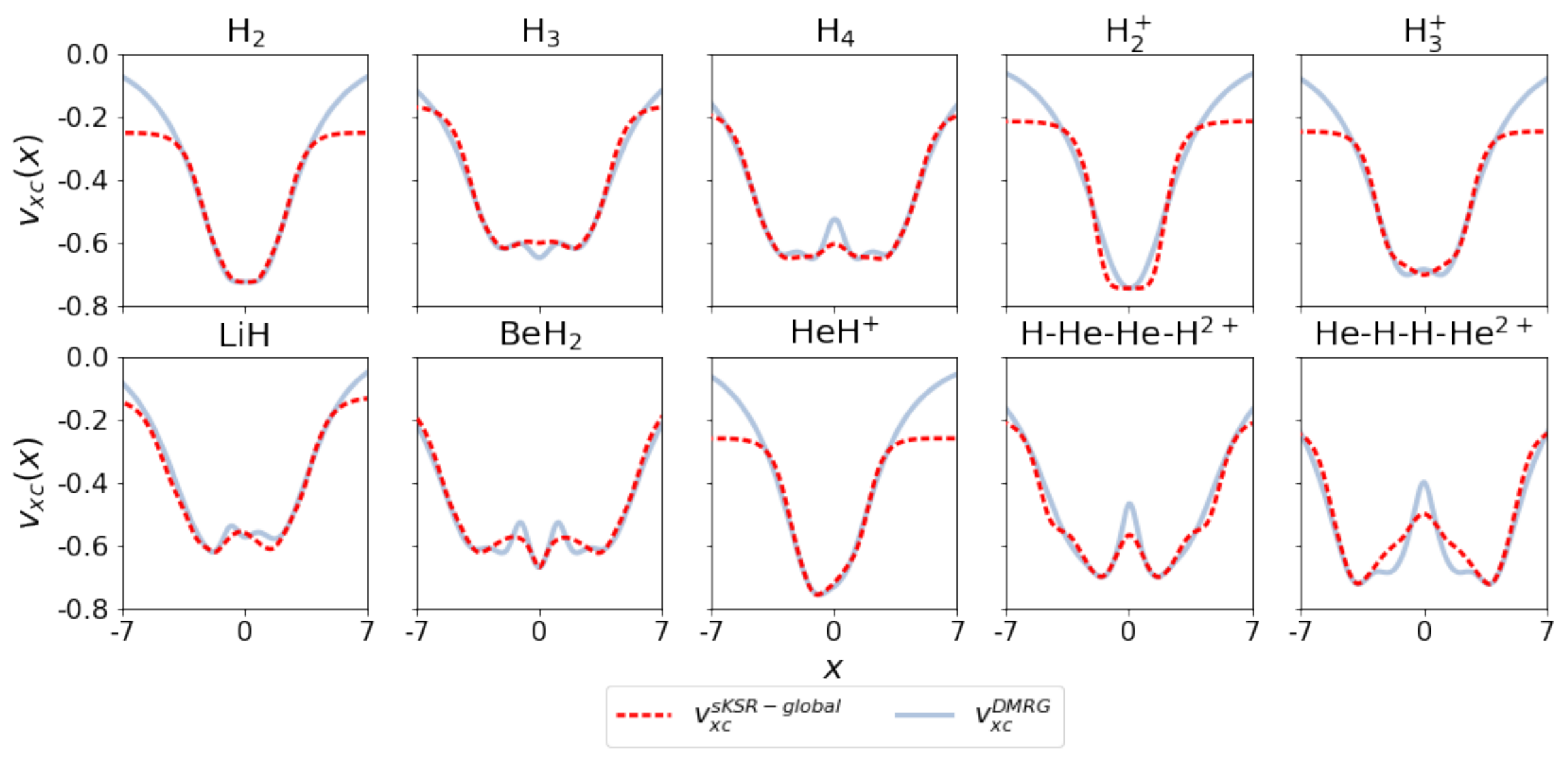}\label{fig:g}}
\caption{(a)  The densities obtained using sKSR-global (orange dashes) and the exact ground-state densities (gray), (b) average XC potentials calculated from sKSR-global approximation (red dashes) to $\epsilon\xc$ and their exact counterparts calculated with DMRG (light blue) for the test molecules in Table.~\ref{tab:datasets} at equilibrium separations. The sKSR potentials are shifted by a constant for a better comparison with the exact XC potentials. sKSR-global was trained on H, He, Li, Be, and Be$^{++}$ and validated on Be$^+$. Note that, in general, these 1D densities and XC potentials can differ even qualitatively from their 3D analogs.}
\label{fig:all_eq_density_vxc}
\end{figure*}

Using sKSR-global, the predicted densities of each molecule have little noticeable error, see Fig.~\ref{fig:all_eq_density_vxc}(a). The corresponding XC potentials are shown in Fig.~\ref{fig:all_eq_density_vxc}(b). For all unpolarized systems, we run restricted KS calculations, and the up and down XC potentials match, while for polarized systems (Li, Be$^+$, H$_2^+$, and H$_3$ only) we run unrestricted KS calculations. The sKSR-LDA and sKSR-GGA total densities and XC potentials for the test set are included in the Supporting Information.
The comparison to exact XC potentials is not expected to be as precise as potentials are extremely sensitive to densities. However, for each of these examples, we see that the sKSR-global XC potential closely mimics the exact XC potential, even though we did not include XC potentials in the training. Furthermore, seemingly large deviations in the XC potentials can result in similar resulting densities. For example, this can be seen in the case of BeH$_2$ where the XC potentials are noticeably different but the resulting densities are very similar. The KS potentials are reasonably accurate for the test set (see Supporting Information). Note that similar to the exact XC potentials, the sKSR-global XC potentials are smooth, due to the use of a smooth activation function.

We can use these potentials to validate the known theoretical properties of the exact XC potentials for different test systems, compare with other XC approximations, and utilize them to introduce corrections to existing local and semilocal approximations. Similarly, sKSR-global can also produce quite accurate spin-densities even though we did not incorporate spin-densities in the loss function while training the XC functionals (see Fig.~A1 in Supporting Information).

\begin{figure}[htp!]
    \centering
    \includegraphics[width= 0.5\textwidth]{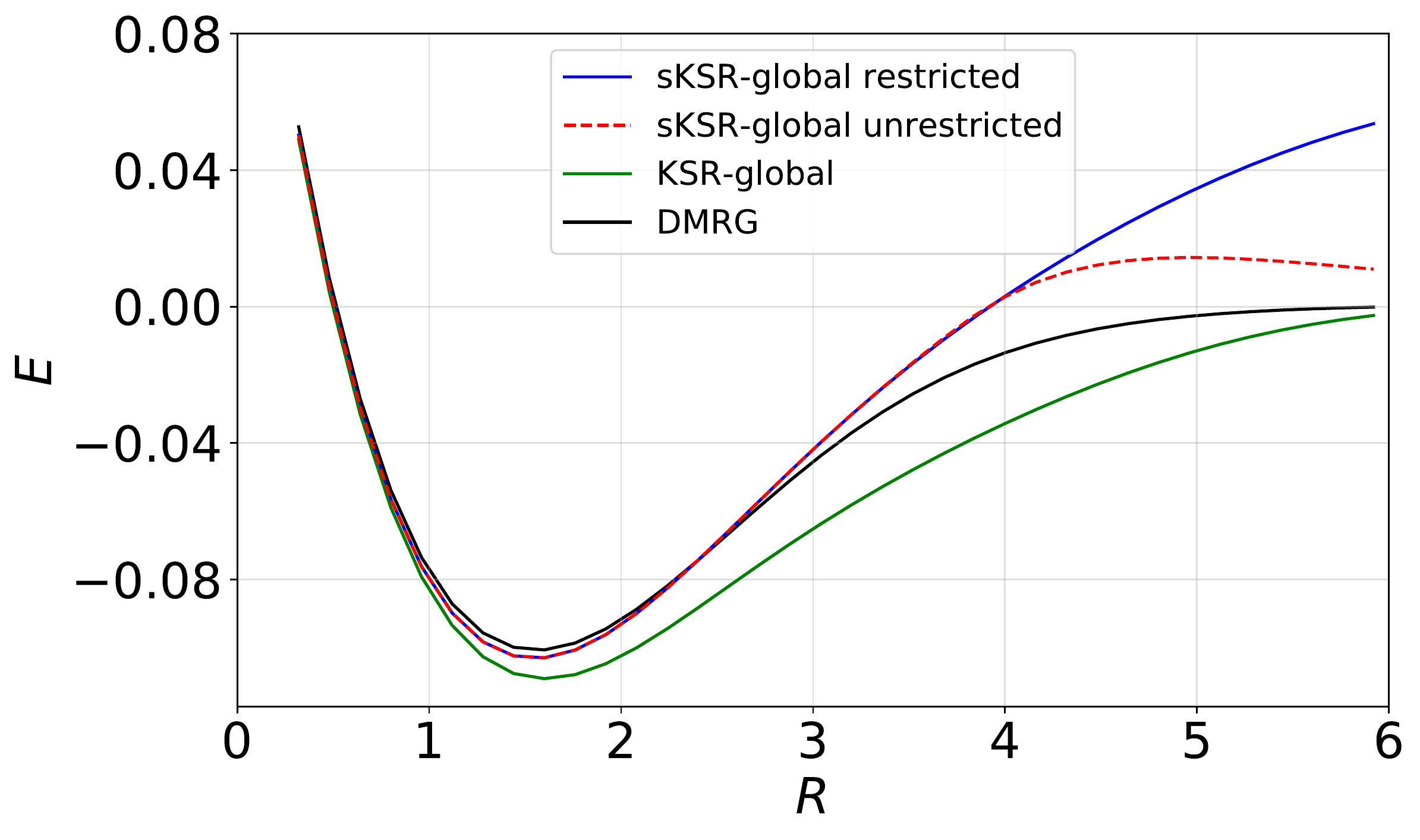}
    \caption{ The binding energy curve of H$_2$ molecule calculated based on the total energy prediction for H$_2$ molecule and the energy of the individual H atoms. sKSR-global was evaluated using restricted KS (blue) and unrestricted KS (red dashes) scheme. The DMRG (black) and KSR-global (green) results are also shown. All the neural approximations, with and without spin, are trained on the dataset given in Table.~\ref{tab:datasets}.} 
   \label{fig:h2_binding_energy_curve}
\end{figure}

 A very interesting question is:
how does our weakly-correlated sKSR behave for strongly-correlated systems?
We answer this by studying the paradigm case of the H$_2$ binding curve in
Fig.~\ref{fig:h2_binding_energy_curve}, where the sKSR-global curve remains highly accurate up to at least
3 Bohr.   Just as with all single-particle methods, the restricted calculation yields energy that is far too high in the
dissociated limit.  On the other hand, an unrestricted calculation, which breaks spin-symmetry beyond about 4 Bohr, does dissociate correctly, but at the price of poor spin densities and a kink in the binding energy
curve.  Fig.~A6 in Supporting Information shows analogous features for sKSR-LDA and sKSR-GGA, and also shows the accuracy of the total
density of the unrestricted solutions at large separations.  
Fig.~\ref{fig:h2_binding_energy_curve} also shows the result of a KSR-global calculation (i.e., total
density only), but trained just on atoms.   While it naturally dissociates correctly, it is much less
accurate.  Of course, the KSR-global of Ref.~\citenum{LHPB21} is chemically accurate for the entire curve because
its training included a stretched bond.

\begin{figure}[htp!]
    \centering
    \includegraphics[width= 0.5\textwidth]{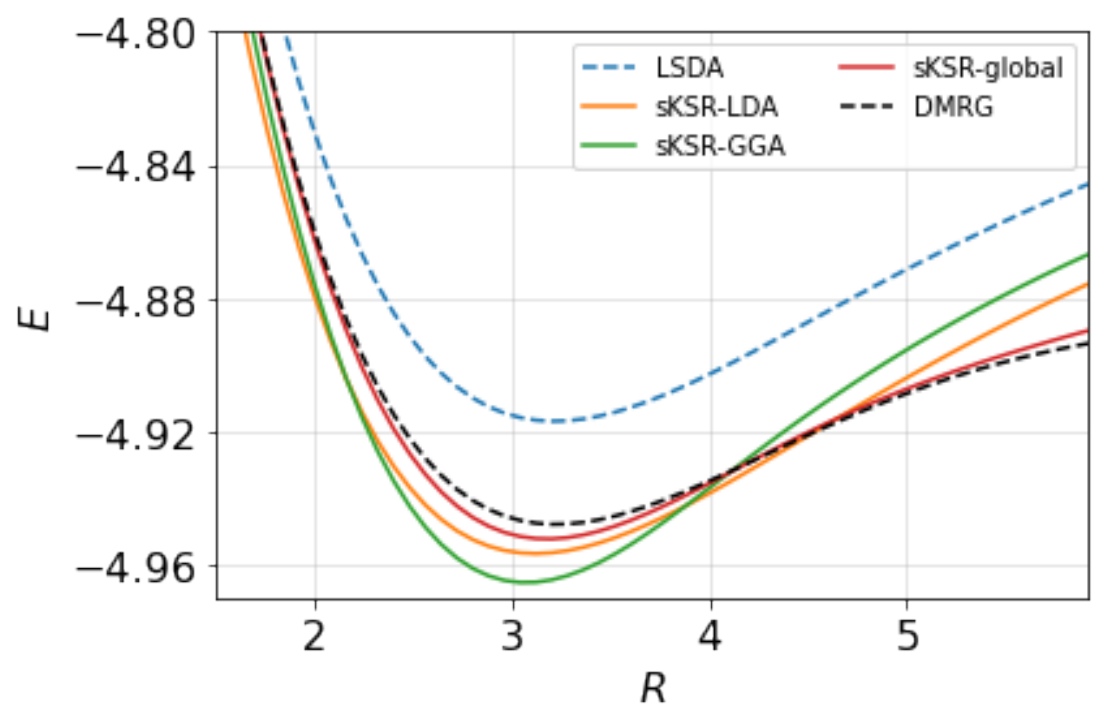}
    \caption{The complete dissociation energy curve of LiH molecule generated with  sKSR-LDA (orange), sKSR-GGA (green) and sKSR-global(red). The DMRG (black dashes) and the uniform gas LSDA (blue dashes) results are also shown. The neural XC functional approximations were trained and validated on atoms and ions given in Table.~\ref{tab:datasets}. }
   \label{fig:lih_binding_energy_curve}
\end{figure}

\begin{figure}[htp!]
     \centering
     \subfloat[]{\includegraphics[width=0.45\textwidth]{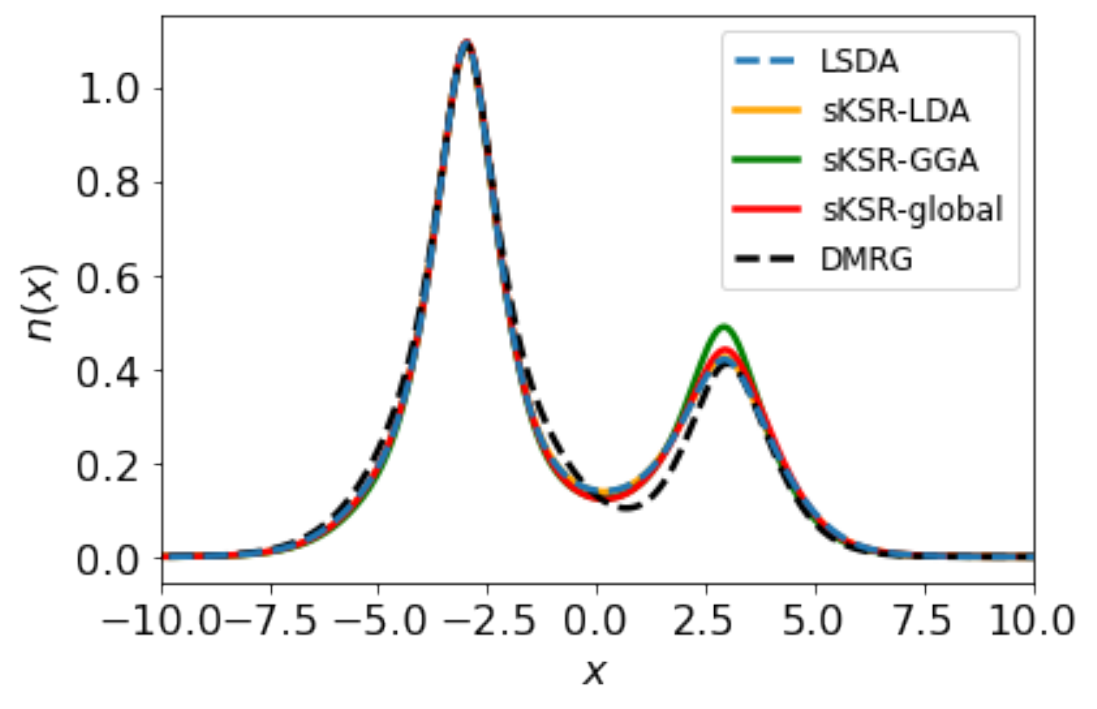}\label{fig:k}}
    \hfill
    \subfloat[]{\includegraphics[width=0.45\textwidth]{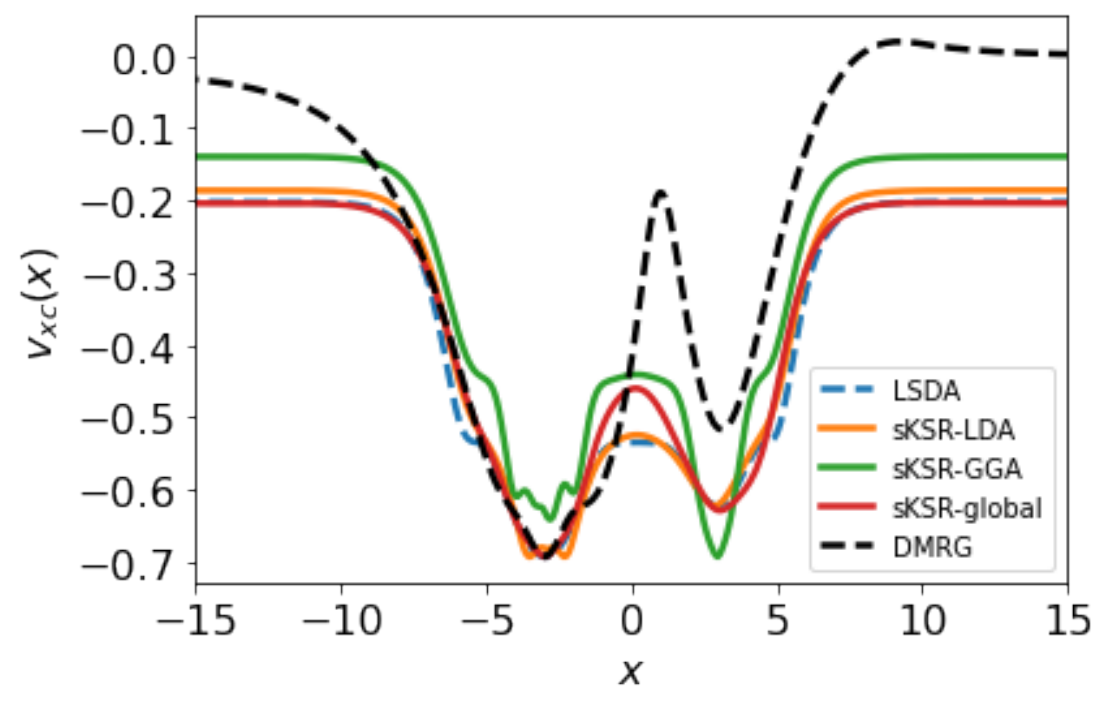}\label{fig:l}}
\caption{ (a) The total density and (b) the average XC potentials of LiH at a bond-distance of 5.92 Bohr calculated with the three neural XC functionals as well as uniform-gas LSDA. The exact (DMRG) average XC potentials are included for comparison.}
\label{fig:lih_592}
\end{figure}

 In many cases, the predictability of sKSR can extend well beyond the equilibrium bond distance. Fig.~\ref{fig:lih_binding_energy_curve} shows the complete dissociation energy curve of LiH obtained from restricted calculation. Near equilibrium, sKSR-LDA and sKSR-GGA underestimate the binding energy but perform better than LSDA. As the bond is stretched, sKSR-GGA and sKSR-LDA quickly deviate from the expected trajectory. However, sKSR-global performs well throughout, extending its predictive accuracy well beyond the equilibrium bond distance. We show the total density and the XC potential of stretched LiH at 5.92 Bohr in Fig.~\ref{fig:lih_592}. LSDA largely overestimates the total energy of the stretched molecule, but its density remains reasonably accurate. The XC potentials calculated from neural XC functional approximations are comparable, with sKSR-global closely approximating the exact behavior. A comparison of the sKSR-global and the exact total density and XC potential of stretched LiH with respect to the atomic contributions from Lithium and Hydrogen is included in the Supporting Information.

The approximate total energy of a molecule can have two types of error contributions: the error due to the approximate functional and the error arising from the self-consistent density~\cite{VKBSS19}.  For most XC functionals, the total density calculated from the self-consistent solution of the KS equations works as an excellent approximation to the exact density for most systems. Hence, the density-driven error is often negligible. However, some approximations can have significant density-driven errors~\cite{KSSB19}. For our test molecules, the errors in the self-consistent densities were trivial and consequently had minimal impacts on the atomization energy errors. The functional and density-driven errors in our neural XC functional approximations are reported for the hydrogen molecule in the supplementary information section.

 We found that sKSR-global achieves remarkable accuracy and generalization in a very data-efficient manner by including the self-consistent KS equations into the training. sKSR-global predicts the ground-state energy of ten unseen molecules in equilibrium with a mean absolute error of 2.7 mH ($\sim$1.7 kcal/mol) when trained with just five atoms and ions.  Hence, a nonlocal XC functional approximation trained on atomic energies and densities can generate predictions for weakly-correlated molecules with near chemical accuracy. An extension of this work can lead to an ML functional that is applicable across a broad chemical spectrum without using an exceedingly large training set. The end-to-end differentiable implementation also ensures smooth and reasonable XC potentials. In addition, sKSR-global trained on atoms can adequately describe a molecule with a stretched bond. Combining differentiable programming with inherent physical intuition thus takes us one step closer to a generalizable, chemically accurate ML XC functional.

The application of the current sKSR algorithm is limited to 1D systems and our test set does not include real 3D molecules. However, the methods presented are transferable to 3D and we anticipate that the characteristic performance is not unique to 1D systems, as these systems tend to mimic their 3D analogs~\cite{WSBW12}. The low-dimensional examples are useful for quick and rigorous assessment of the quality of an approximation. Besides, the predictions from the local and semilocal approximation explored in our study are consistent with the 3D differentiable formulations in Ref~\citenum{KV21} and Ref.~\citenum{DS21}.  

\begin{acknowledgments}
This work is supported by National Science Foundation, grant no. DGE-1633631 (B. K.), CHE-1856165 (B. K., K. B.), and Department of Energy, grant no. DE-SC0008696 (R. P.).
\end{acknowledgments}

\section{Supporting Information}
The Supporting Information is available at \href{https://pubs.acs.org/doi/10.1021/acs.jpclett.2c00371}{https://pubs.acs.org/doi/10.1021/acs.jpclett.2c00371.}.

\section{Data Availability Statement}
The training and testing data and the one-dimensional density functional theory 
solver used for the uniform electron-gas LDA calculations are available at \href{https://github.com/pedersor/DFT_1d}{https://github.com/pedersor/DFT$\_$1d}. The ML models and the JAX version of the DFT code are available from the corresponding author upon reasonable request.

\clearpage

\bibliography{master}

\end{document}